\documentclass[prd,reprint,showpacs,showkeys]{revtex4}
\usepackage{amsfonts}
\usepackage{amssymb}
\usepackage{amsmath}
\usepackage{graphicx}
\usepackage{float}
\usepackage[font={footnotesize,it}]{caption}

\begin{document}

\title{Fate of a Thin-Shell Wormhole Powered by Morris-Thorne Wormhole}
\author{S. Danial Forghani}
\email{danial.forghani@emu.edu.tr}
\author{S. Habib Mazharimousavi}
\email{habib.mazhari@emu.edu.tr}
\author{M. Halilsoy}
\email{mustafa.halilsoy@emu.edu.tr}
\affiliation{Department of Physics, Faculty of Arts and Sciences, Eastern Mediterranean
University, Famagusta, North Cyprus via Mersin 10, Turkey}

\begin{abstract}
Asymmetric thin-shell wormholes from two traversable Morris-Thorne wormhole
spacetimes, with identical shape but different redshift functions, are
constructed. Energy density of the thin-shell wormhole derives its power
from a Morris-Thorne wormhole which is already exotic. By choice, the weak
energy condition for the thin-shell wormhole is satisfied. A linear
barotropic equation of state is assumed to hold after the radial
perturbations. The fate of our thin-shell wormhole, after the perturbation,
is striking: the asymmetric thin-shell wormhole is destined either to
collapse to the original Morris-Thorne wormhole or expand indefinitely along
with the radius of the throat. In case it collapses to the original
wormhole, the result is an asymmetric Morris-Thorne wormhole. Although this
asymmetry does not reflect into the embedding diagram of the wormhole,
passing across the throat, the wormhole adventurer feels a different
redshift function.
\end{abstract}

\maketitle

\section{Introduction}

The project of minimizing the indispensable exotic matter necessary in
constructing wormholes was accomplished successfully under the title of
thin-shell wormholes (TSWs) \cite{Visser}. The idea was to confine the
entire notorious negative energy density to a narrow band, ideally at the
scales of Planck length. The reason in doing this is to earn legitimacy to
such a negative energy concept within the domain of classical physics. The
cut and paste procedure may cover anything to anything provided the
consistency requirement of junction conditions are met. The expanding
literature on TSWs is the best evidence to support the proposal of Visser.
Furthermore, the idea is not restricted by spherical symmetry above but
equally well finds rooms of application in cylindrical symmetry, albeit with
more strict regulations.

Besides symmetric TSWs, which has been employed in the past \cite%
{Collective1}, recently we have promoted the idea of asymmetric thin-shell
wormholes (ATSWs) which hosts non-isotropic and inhomogeneous spacetimes 
\cite{Forghani1, Forghani2}. Thus, when Alice crosses the throat into the
other universe, she encounters a truly different world.

On the other hand, there exists in the wormhole literature spacetimes which
represent wormholes in essence \cite{Collective2}. As explained in detail by
Morris and Thorne \cite{Morris}, asymptotic flatness, flare-out conditions
and other conditions must all be satisfied for such wormhole spacetimes in
order to have a safe wormhole travel across them. Expectedly, some of these
conditions such as asymptotic flatness must be taken in a more restrictive
sense when one deals with cylindrical symmetries instead of spherical.
However, conditions like asymptotic flatness in radial infinity when one
concerns the throat's neighborhood only \cite{Bronnikov}, or flare-out
condition when it comes to TSWs \cite{Mazhari}, could be disregarded
unrestrictedly.

Our aim in this paper is to construct a TSW from two different patches of
Morris-Thorne traversable wormholes (MTWs). The latter is characterized by
two basic functions and a key parameter; these are respectively the redshift
function $\Phi \left( r\right) $, the shape function $b\left( r\right) $,
and the minimum radius $r_{0}$ measuring the throat. We glue two such MTWs
with different $\Phi \left( r\right) $\ functions to construct a viable
ATSW. Our ATSW will naturally draw its energy from MTW, which will be
exotic. However, degree of freedom available at our disposal allows us to
exploit the choice of throat radius of our ATSW such that the weak energy
condition (WEC), i.e. $\sigma \geq 0$ and $\sigma +p\geq 0$, holds under
some certain conditions, where $\sigma $\ and $p$\ refer to the energy
density and the tangential pressure on the ATSW's throat, respectively. This
takes care of exotic energy problem in TSWs, whereas another serious problem
concerns about the stability. To work out the latter, we introduce a
barotropic fluid equation of state (EoS) after the perturbation and
investigate the consequences. Based on such a barotropic fluid, we conclude
some results in the following sections.

The organization of the paper is as follows. In section $II$\ we give a
general description of ATSW construction from MTW. Stability analysis of our
ATSW is considered in section $III$. The paper is completed with conclusion
in part $IV$.

\section{ATSWs from wormhole spacetimes}

In this section we begin by setting a general $4-$dimensional spacetime
metric in spherical coordinates%
\begin{equation}
ds^{2}=-A\left( \ell \right) dt^{2}+B\left( \ell \right) d\ell ^{2}+C\left(
\ell \right) d\Omega ^{2},
\end{equation}%
in which the metric functions $A\left( \ell \right) $, $B\left( \ell \right) 
$ and $C\left( \ell \right) $\ are all positive functions of radial
coordinate $\ell $, whereas $d\Omega ^{2}$ is traditionally a $2-$sphere
line element defined by $d\Omega ^{2}=d\theta ^{2}+\sin ^{2}\theta d\varphi
^{2}$. On this spacetime we define a time-dependent timelike hypersurface
expressed by%
\begin{equation}
\mathcal{H}\left( \ell ,\tau \right) :=\ell -a\left( \tau \right) =0,
\end{equation}%
in which $\tau $\ is the proper time measured by the observer defined
through the constraint $-A\left( \ell \right) dt^{2}+B\left( \ell \right)
d\ell ^{2}=-d\tau ^{2}$ on the hypersurface. Therefore, the induced metric
on the hypersurface $\mathcal{H}$%
\begin{equation}
ds_{\mathcal{H}}^{2}=-d\tau ^{2}+C\left( a\right) d\Omega ^{2}
\end{equation}%
admits%
\begin{equation}
\dot{t}^{2}=\frac{1+B\dot{a}^{2}}{A},
\end{equation}%
where an overdot stands for a total derivative with respect to the proper
time $\tau $. Then, the spacelike normal $4-$vector components to the
hypersurface are denoted by%
\begin{equation}
n_{\mu }=\left( g^{\alpha \beta }\frac{\partial \mathcal{H}}{\partial
x^{\alpha }}\frac{\partial \mathcal{H}}{\partial x^{\beta }}\right) ^{-1/2}%
\frac{\partial \mathcal{H}}{\partial x^{\mu }},
\end{equation}%
which can be used to attain the components of the $3\times 3$ curvature
tensor on $\mathcal{H}$ as follows%
\begin{equation}
K_{ab}=-n_{\mu }\left( \frac{\partial x^{\mu }}{\partial \xi ^{a}\partial
\xi ^{b}}+\Gamma _{\alpha \beta }^{\mu }\frac{\partial x^{\alpha }}{\partial
\xi ^{a}}\frac{\partial x^{\beta }}{\partial \xi ^{b}}\right) .
\end{equation}%
Herein, the Christoffel symbols $\Gamma _{\alpha \beta }^{\mu }$ are the
ones compatible with the metric of the bulk spacetime $g_{\alpha \beta }$,
while $x^{\alpha }=\left\{ t,\ell ,\theta ,\varphi \right\} $ and $\xi
^{a}=\left\{ \tau ,\theta ,\varphi \right\} $ stand for the coordinates of
the bulk spacetime and the hypersurface, respectively. Using Eqs. (3)-(6)
one acquires the explicit form of the mixed extrinsic curvature tensor given
by%
\begin{equation}
K_{b}^{a}=diag\left( K_{\tau }^{\tau },K_{\theta }^{\theta },K_{\varphi
}^{\varphi }\right) =\sqrt{\frac{B}{1+B\dot{a}^{2}}}diag\left[ \ddot{a}+%
\frac{A^{\prime }}{2AB}+\frac{\dot{a}^{2}}{2}\left( \frac{A^{\prime }}{A}+%
\frac{B^{\prime }}{B}\right) ,\frac{1+B\dot{a}^{2}}{2B}\left( \frac{%
C^{\prime }}{C}\right) ,\frac{1+B\dot{a}^{2}}{2B}\left( \frac{C^{\prime }}{C}%
\right) \right] ,
\end{equation}%
with its trace%
\begin{equation}
K\equiv K_{a}^{a}=\sqrt{\frac{B}{1+B\dot{a}^{2}}}\left[ \ddot{a}+\frac{1}{2B}%
\left( \frac{A^{\prime }}{A}+2\frac{C^{\prime }}{C}\right) +\frac{\dot{a}^{2}%
}{2}\left( \frac{A^{\prime }}{A}+\frac{B^{\prime }}{B}+2\frac{C^{\prime }}{C}%
\right) \right] .
\end{equation}%
In the last two equations a prime represents a total derivative with respect
to the metric functions' argument $\ell $.

Having calculated all these, the following is how to construct a TSW in
Visser's sense \cite{Visser}. Assume we have two (not necessarily the same)
spacetimes of the general metric brought in Eq. (1). From both spacetimes,
the inner part of the hypersurface $\mathcal{H}$\ gets cut out and the outer
parts are glued together at $\mathcal{H}$, which now becomes the two
spacetimes' $2+1-$dimensional boundary; the whole structure is called a TSW
and the hypersurface $\mathcal{H}$\ becomes the throat of the wormhole. We
note that the hypersurface is selected such that the cut radius $a$\ is
greater than any possible horizon existed in the two spacetimes. If the
metric functions $A\left( \ell \right) $ and $B\left( \ell \right) $ of the
two spacetimes are exactly the same, the result is a symmetric TSW, while if
they are not (under certain circumstances),\ an ATSW \cite{Forghani1}.

According to Lanczos \cite{Lanczos}\ and Israel \cite{Israel}, such a
structure must follow two certain conditions. The first condition, which was
implicitly thought of in the explanations above, states that the metric of
the throat must be continuous across the shell / throat i.e., $\left[ g_{ij}%
\right] =g_{ij}^{\left( 2\right) }-g_{ij}^{\left( 1\right) }=0$. This
implies that the metric function $A\left( \ell \right) $, $B\left( \ell
\right) $ and $C(\ell )$ should satisfy%
\begin{equation}
\left\{ 
\begin{array}{c}
-A_{1}\left( a\right) \dot{t}_{1}^{2}+B_{1}\left( a\right) \dot{a}%
^{2}=-A_{2}\left( a\right) \dot{t}_{2}^{2}+B_{2}\left( a\right) \dot{a}%
^{2}=-1 \\ 
C_{1}\left( a\right) =C_{2}\left( a\right) =C\left( a\right)%
\end{array}%
\right.
\end{equation}%
on the throat, where the indices $1$ and $2$ distinguish the bulk spacetimes.

Meanwhile, the second condition, imposes a discontinuity on the curvature
tensor $K_{b}^{a}$ passing across the throat, and relates it to the
energy-momentum tensor $S_{b}^{a}$($=diag\left( -\sigma ,p,p\right) $) of
the matter that unavoidably must exist on the throat. This condition is
mathematically expressed as ($8\pi G=1$)%
\begin{equation}
\left[ K_{b}^{a}\right] -\delta _{b}^{a}\left[ K\right] =-S_{b}^{a},
\end{equation}%
where the square brackets impart a jump in the included entity i.e. $\left[
\Upsilon \right] =\Upsilon _{2}-\Upsilon _{1}$. The latter equation,
together with Eqs. (7) and (8) amount to two equations for the energy
density $\sigma $\ and the angular pressure $p$\ on the throat as follow%
\begin{equation}
\sigma =-\frac{C^{\prime }}{C}\sum_{i=1}^{2}\sqrt{\frac{1+B_{i}\dot{a}^{2}}{%
B_{i}}}
\end{equation}%
and%
\begin{equation}
p=\sum_{i=1}^{2}\left\{ \sqrt{\frac{B_{i}}{1+B_{i}\dot{a}^{2}}}\left[ \ddot{a%
}+\frac{\left( A_{i}B_{i}C\right) ^{\prime }}{2A_{i}B_{i}C}\dot{a}^{2}+\frac{%
\left( A_{i}C\right) ^{\prime }}{2B_{i}A_{i}C}\right] \right\} .
\end{equation}%
By manipulating Eq. (11) one is able to extract a dynamic mechanical
equation for the throat such as%
\begin{equation}
\dot{a}^{2}+V\left( a\right) =0,
\end{equation}%
where the potential%
\begin{equation}
V\left( a\right) =\frac{1}{2}\left( \frac{1}{B_{1}}+\frac{1}{B_{2}}\right) -%
\left[ \frac{C^{\prime }}{2C\sigma }\left( \frac{1}{B_{1}}-\frac{1}{B_{2}}%
\right) \right] ^{2}-\left( \frac{C\sigma }{2C^{\prime }}\right) ^{2}
\end{equation}%
against the radius $a$\ determines the stability of the throat under a
radial perturbation. Note that in case of a perfect similarity between the
two bulk spacetimes($A_{1}\left( a\right) =A_{2}\left( a\right) $ and $%
B_{1}\left( a\right) =B_{2}\left( a\right) $), the equations above would
simplify to a great deal.

Before we proceed, let us be more specific about the metric of the two
spacetimes. The metric of the MTW \cite{Morris}\ is traditionally expressed
in the form%
\begin{equation}
ds_{MT}^{2}=-e^{2\Phi \left( r\right) }dt^{2}+\frac{dr^{2}}{1-\frac{b\left(
r\right) }{r}}+r^{2}d\Omega ^{2},
\end{equation}%
in which $\Phi \left( r\right) $ and $b\left( r\right) $ are known as the
redshift and the shape functions, respectively. For the spacetime to
represent a proper spatial wormhole geometry and not to have any horizons or
singularities, these two functions obey the following conditions; throughout
the spacetime $1-b\left( r\right) /r\geq 0$, and $\Phi \left( r\right) $\ is
finite everywhere. Therefore, the wormhole possesses a throat at $%
r_{0}=b(r_{0})$ as the minimum radius.

One example of such functions which fits to the criterion above for the
shape function is%
\begin{equation}
b\left( r\right) =\frac{r_{0}^{2}}{r},
\end{equation}%
with which, alongside the new radial coordinate%
\begin{equation}
\ell =\pm \left( r^{2}-r_{0}^{2}\right) ^{1/2},\text{ \ \ }-\infty <\ell
<\infty
\end{equation}%
the MTW metric transforms to%
\begin{equation}
ds_{MT}^{2}=-e^{2\Phi \left( \ell \right) }dt^{2}+d\ell ^{2}+(\ell
^{2}+r_{0}^{2})d\Omega ^{2}.
\end{equation}

Over the years, while many spacetimes have been studied in the TSW
framework, the MT spacetime has amazingly been ignored; perhaps because the
idea of constructing a TSW out of a traversable wormhole is bizarre enough
not to draw attentions. Nonetheless, we dare to use the metric in Eq. (15)
to construct a TSW, though, not even a symmetric one but an ATSW by
entailing different redshift functions on the two sides, i.e. $\Phi
_{1}\left( \ell \right) \neq \Phi _{2}\left( \ell \right) $. Therefore, the
energy density $\sigma $, the angular pressure $p$, and the potential $%
V\left( a\right) $ of the throat versus its radial coordinate amount to ($%
\ell =a$)%
\begin{equation}
\sigma =-\frac{4a\sqrt{1+\dot{a}^{2}}}{(a^{2}+r_{0}^{2})},
\end{equation}%
\begin{equation}
p=\sqrt{1+\dot{a}^{2}}\left( \frac{2\ddot{a}}{1+\dot{a}^{2}}+\Phi
_{1}^{\prime }+\Phi _{2}^{\prime }+\frac{2a}{a^{2}+r_{0}^{2}}\right) ,
\end{equation}%
and%
\begin{equation}
V\left( a\right) =1-\left[ \frac{\left( a^{2}+r_{0}^{2}\right) \sigma }{4a}%
\right] ^{2},
\end{equation}%
respectively. Let us add that herein $a$ is the radius of the throat of the
ATSW while $r_{0}$ stands for the radius of the original MTW. Hence, it
holds that $a\geq 0$.

In the case of a static ATSW (not necessarily at equilibrium) one sets $%
a=a_{0}$ and $\dot{a}=\ddot{a}=0$ which in turn imply%
\begin{equation}
\sigma _{0}=-\frac{4a_{0}}{(a_{0}^{2}+r_{0}^{2})},
\end{equation}%
and%
\begin{equation}
p_{0}=\Phi _{1}^{\prime }\left( a_{0}\right) +\Phi _{2}^{\prime }\left(
a_{0}\right) +\frac{2a_{0}}{a_{0}^{2}+r_{0}^{2}}.
\end{equation}%
As we mentioned above, generally $a_{0}\geq 0$ but for the specific case
when $a_{0}=0$ the two throats (throat of the original MTW and the throat of
the ATSW) coincide; It is like cutting two bulk wormhole metrics at their
throats and rejoining them at ATSW's throat. Under these circumstances, $%
\sigma _{0}=0$, which implies that there is no need of matter at the throat.
On the other hand, $p_{0}=\Phi _{1}^{\prime }\left( a_{0}\right) +\Phi
_{2}^{\prime }\left( a_{0}\right) $, which is generally nonzero and could
potentially be positive. This entails that, under static condition, the weak
and so the null energy conditions are satisfied at the thin-shell's throat.
Furthermore, it is worth mentioning that setting $a_{0}=0$ actually turns
the ATSW to an asymmetric wormhole. A specific choice of $\Phi _{1}=\Phi
_{2}=\Phi $ turns the resultant ATSW into a good old TSW. For the latter
case one finds $p_{0}=2\Phi ^{\prime }\left( 0\right) .$

\section{Stability analysis of MT powered ATSWs}

With reference to Eq. (21), in this section we would like to further analyze
the potential of such a structure by employing a linear barotropic EoS, i.e.%
\begin{equation}
p=p_{0}+\omega \left( \sigma -\sigma _{0}\right) ,
\end{equation}%
wherein $\sigma _{0}$ and $p_{0}\ $are given in Eqs. (22) and (23), and $%
\omega $ is the propagation speed of sound through the matter on the throat;
accordingly, $\omega \in \left( 0,1\right) $. This form of EoS guarantees
that when $a\rightarrow a_{0}$ we get $p\rightarrow p_{0}$.

Besides, applying the energy conservation on the throat, $\nabla _{j}S^{ij}=0
$, when $i=\tau $, one is able to recover the equation%
\begin{equation}
\sigma ^{\prime }+\frac{2a}{a^{2}+r_{0}^{2}}\left( \sigma +p\right) =0
\end{equation}%
for the energy density and the tangential pressure on the throat. Using the
above equation together with the EoS in Eq. (24), we arrive at an expression
for the energy density 
\begin{equation}
\sigma =\frac{1}{\omega +1}\left[ \left( \sigma _{0}+p_{0}\right) \left( 
\frac{a_{0}^{2}+r_{0}^{2}}{a^{2}+r_{0}^{2}}\right) ^{\omega +1}+\omega
\sigma _{0}-p_{0}\right] ,
\end{equation}%
which with Eqs. (22) and (23), can be used to rewrite the potential at Eq.
(21) as%
\begin{equation}
V\left( a\right) =1-\left\{ \frac{\left( a^{2}+r_{0}^{2}\right) }{4a\left(
\omega +1\right) }\left[ \left( \Psi -\frac{2a_{0}}{a_{0}^{2}+r_{0}^{2}}%
\right) \left( \left( \frac{a_{0}^{2}+r_{0}^{2}}{a^{2}+r_{0}^{2}}\right)
^{\omega +1}-1\right) -\frac{4a_{0}}{a_{0}^{2}+r_{0}^{2}}\left( \omega
+1\right) \right] \right\} ^{2},
\end{equation}%
where $\Psi =$\ $\Phi _{10}^{\prime }+\Phi _{20}^{\prime }$ is a constant.
Since $r_{0}$ is merely a scale factor, without\ loss of generality we can
set it equal to $1$. Having this done, out of $V\left( a\right) $, taking
the first derivative with respect to $a$ and setting $a=a_{0}$, we observe
that $V^{\prime }\left( a_{0}\right) $ has an extremum at a critical radius\
given by 
\begin{equation}
a_{c}=\frac{\left( \sqrt{27+\Psi ^{2}}+\sqrt{27}\right) ^{2/3}-\Psi ^{2/3}}{%
\sqrt{3}\left[ \Psi \left( \sqrt{27+\Psi ^{2}}+\sqrt{27}\right) \right]
^{1/3}},
\end{equation}%
which is independent of $\omega $. Our numerical calculations (Fig. 1)
confirms that the second derivative of the potential at the critical radius $%
a_{C}$ i.e. $V^{\prime \prime }\left( a_{C}\right) $ is negative definite
for admissible domain of $\omega $ and arbitrary values of $\Psi $.
Therefore, $a_{C}$ represents an unstable equilibrium radius for the TSW. On
the other hand, if one rewrites Eq. (28) for $\Psi $ to obtain%
\begin{equation}
\Psi =\frac{2}{a_{C}\left( a_{C}^{2}+1\right) },
\end{equation}%
\begin{figure}[tbp]
\includegraphics[width=80mm,scale=0.7]{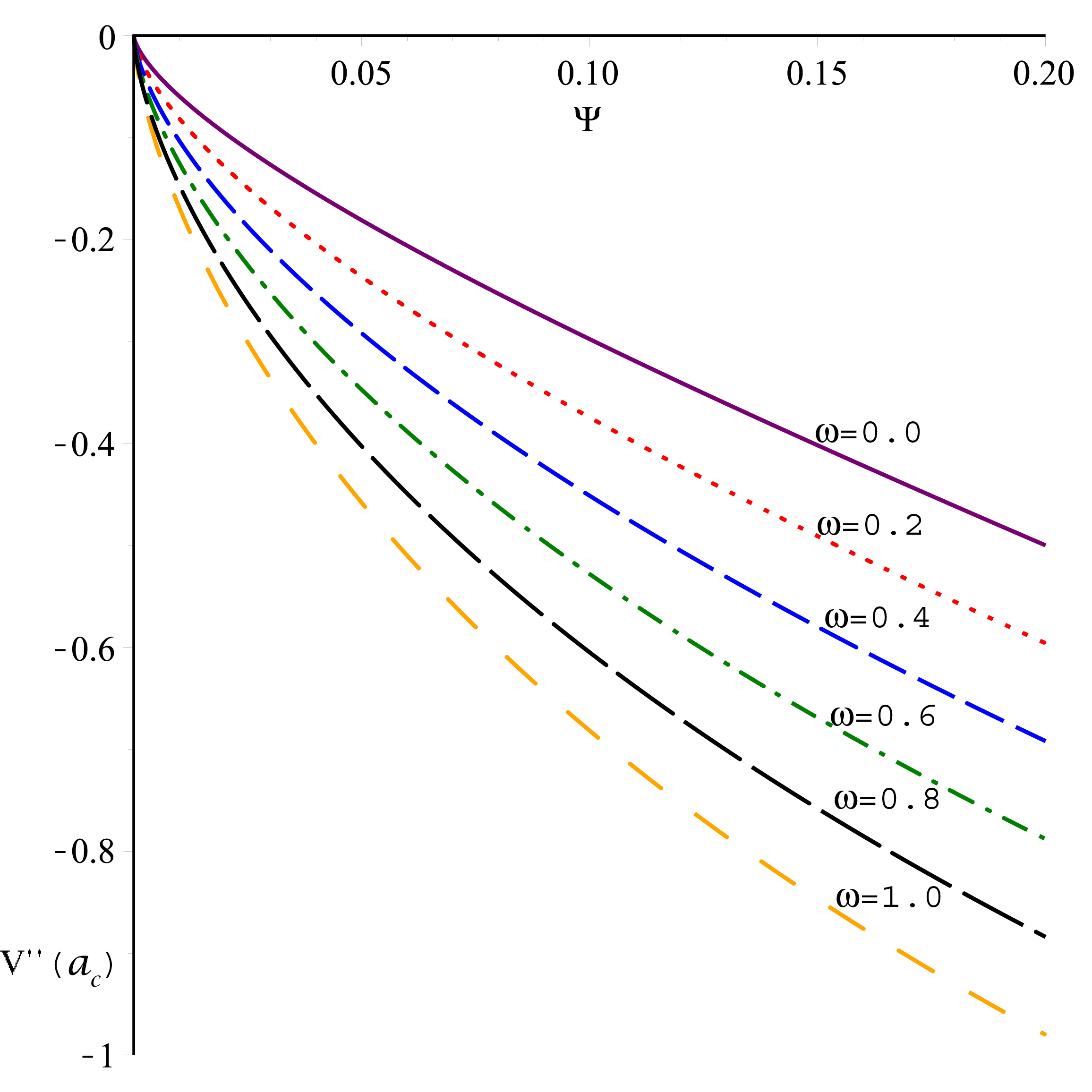}
\caption{{}$V^{\prime \prime }\left( a_{C}\right) $ versus $\Psi $ for
various values of $\protect\omega $\ within its permitted range. The figure
accents that $V^{\prime \prime }\left( a_{C}\right) $\ is negative definite,
saying that $a_{C}$ is the maximum point of the potential $V\left( a\right) $%
.}
\end{figure}
then it could be brought this back to $V^{\prime }\left( a_{0}\right) $ to
see that for any $a_{0}<a_{C}$ the numerical value of $V^{\prime }\left(
a_{0}\right) $ is positive, while for any $a_{0}>a_{C}$ this value is
negative. Accordingly, we state that, if the two MTW spacetimes are glued
together at a radius $a_{0}<a_{C}$, the TSW collapses to the original
wormhole ($V\left( a\right) \rightarrow -\infty $ as $a\rightarrow 0$),
while, if $a_{0}$ is selected such that $a_{0}>a_{C}$, the TSW evaporates ($%
V\left( a\right) \rightarrow -\infty $ as $a\rightarrow \infty $). Note
that, if $\Phi _{1}$ and $\Phi _{2}$ were selected such that $\Psi =0$, no
equilibrium point would exist and $V^{\prime }\left( a_{0}\right) $ would
always be positive, stating that the TSW would again collapse to the
original wormhole. In either case of collapsing to the original wormhole,
since there is no restrictions on the redshift functions $\Phi _{1}$ and $%
\Phi _{2}$\ (except for their finiteness throughout the corresponding
spacetimes), the resultant wormhole will be an asymmetric wormhole, in
general. Stated otherwise, $b\left( r\right) $ function is common to both
MTV whereas the redshift functions will differ.

\section{Conclusion}

The first prototype example of a traversable wormhole was introduced in 1988
by Morris and Thorne \cite{Morris}, where they investigated the physical
details of an observer for a safe passage through a wormhole structure.
Exotic matter and inherent instability constitute the perilous items that
threaten such a passage. Employing such MTWs, we construct TSWs, both
symmetric and asymmetric at equal ease. The energy density remains still
negative, unless we adjust the radius of both items to make it zero. The
latter case renders satisfaction of the WEC possible, leaving the pressure
components dependant on the redshift functions. Radial perturbation of the
throat gives rise to an energy-momentum tensor which is identified as the
EoS for a barotropic fluid with a critical radius $a_{C}$. Steep rise
towards/away from the critical radius indicates a collapsing/ever expanding
TSW; either it collapses to the original MTW or attains a negative infinite
potential in the equation $\dot{a}^{2}+V\left( a\right) =0$. Such a behavior
determines the unstable fate of a TSW established on the MTW. It is
worthwhile to note that, since the constructed TSW is asymmetric in general,
in case it collapses to the original MTW, the resultant construction will be
an asymmetric MTW i.e. same shape function but different redshift functions.
Although this asymmetry does not reflect into the embedding diagram of the
wormhole \cite{Morris, Lobo} (since the shape function $b\left( r\right) $\
which determines the geometrical shape of the embedding), due to different
redshift functions considered here, it emanates in the traveler's perception
of time, once they pass the throat. Let's add that asymmetric wormhole in
different framework has been considered in literature \cite{ASW}. It remains
to be seen, whether the discussed ends for other TSWs powered by the MTW is
generic or not. Our choice of the linear barotropic EoS is the simplest one;
will there be different results otherwise? For example, what happens if a
variable EoS \cite{Varela}\ or a Chaplygin gas \cite{Chaplygin, Eiroa}\ are
employed?

\end{document}